\documentclass[pra,onecolumn, showpacs, showkeys, secnumarabic, aps, amsmath, amssymb, nofootinbib, superscriptaddress, longbibliography, floatfix, table-of-contents, dblfloatfix]{revtex4-2}

\usepackage[utf8]{inputenc}
\usepackage[pdftex]{graphicx}
\usepackage{mathrsfs}
\usepackage[colorlinks, breaklinks, urlcolor={blue}, linkcolor={blue}, citecolor={blue}]{hyperref}
\usepackage{array}
\usepackage{amsmath}
\usepackage{type1cm}
\usepackage{graphicx}
\usepackage[english]{babel}
\usepackage{lmodern}
\usepackage{microtype}
\usepackage{booktabs}
\usepackage{caption}
\usepackage{braket}
\usepackage{xcolor}
\usepackage{orcidlink}

\begin{document}
\title{Unitary and non-unitary operators leverage perfect and imperfect single qutrit teleportation}
\author{Sovik Roy \orcidlink{0000-0003-4334-341X} }
\email[]{s.roy2.tmsl@ticollege.org}
\affiliation{Department of Mathematics, Techno Main Salt Lake (Engg. Colg.), \\Techno India Group, EM 4/1, Sector V, Salt Lake, Kolkata  700091, India}
\affiliation{Centre of Advanced Studies and Innovation Lab, 18/27 Kali Mohan Road,
Tarapur, Silchar 788003, INDIA}
\author{Anushree Pandey \orcidlink{0000-0002-0655-1773} }
\email[]{a.bhattacharya.tmsl@ticollege.org}
\affiliation{Department of Mathematics, Techno Main Salt Lake (Engg. Colg.), \\Techno India Group, EM 4/1, Sector V, Salt Lake, Kolkata  700091, India}
\affiliation{Centre of Advanced Studies and Innovation Lab, 18/27 Kali Mohan Road,
Tarapur, Silchar 788003, INDIA}
\author{Tushar Kanti Dey }
\email[]{tkdey54@gmail.com}
\affiliation{Centre of Advanced Studies and Innovation Lab, 18/27 Kali Mohan Road,
Tarapur, Silchar 788003, INDIA}

\author{Surajit Sen  \orcidlink{0000-0003-3537-2173} }
\email[]{ssen55@yahoo.com}
\affiliation{Centre of Advanced Studies and Innovation Lab, 18/27 Kali Mohan Road,
Tarapur, Silchar 788003, INDIA}

\begin{abstract}
\noindent  Teleportation, a novel scheme, initially posited by Bennett \textit{et.al}, has been studied here in the context of sending a single qutrit from Alice to Bob using two qutrit entangled channels as resources. In this paper we have considered two special two qutrit entangled states, which belong to $SU(3)$ group, as useful resources for teleportation. For the successful teleportation, these entangled states have been chosen as quantum channels shared between Alice and Bob. Another entangled basis of two qutrit states have been used as auxiliary states, which would help Alice to manipulate with her channel so that the single qutrit she holds can be successfully teleported to Bob. Bob's choices of measurement operators influence the retrieval of Alice's single qutrit.
\end{abstract}


\maketitle
\section{Introduction:}\label{sec:introduction}
\noindent Ever since Bennett \textit{et.al} proposed their teleportation scheme\cite{bennett1993teleporting} and its subsequent experimental verification done by Bouwmeester et al\cite{bouwmeester1997experimental}, the protocol of teleportation was verified under different physical conditions and with different bipartite and multipartite states as quantum channels\cite{vaidman1994teleportation,davidovich1994teleportation,braunstein1998teleportation,boschi1998experimental,zubairy1998quantum,murao1999quantum,lee2000entanglement,paulson2019tripartite,kumar2013optimal, romano2010teleportation, roa2015probabilistic}. Although dealing with higher dimensional system is a challenging task, the continuation with the study of teleportation schemes in higher-dimensional systems has also not been an exception and many proposals in this context have thus been put forward\cite{pan2006classification,luo2019quantum,fonseca2019high,de2021efficient,roy2014cloned,mei2010probabilistic,garg2021teleportation,tian2011quantum}. Two factors that need to be taken care of in the teleportation schemes are, (a) the channel that sender Alice and receiver Bob share as an entangled resource and (b) the unitary operator or projector that Bob is going to use at his end after receiving the message of the Alice’s measurement outcomes, which Alice conveys to him classically. Bennett \textit{et. al}\cite{bennett1993teleporting, nielsen2010quantum} used one of the Bell states as entangled resources and performed their teleportation scheme by using all the four Bell basis states which they make Alice to manipulate in such a way so that in this local operation and classical communication (LOCC) model, Bob is made to use suitable unitary operator to complete the task and thereby retrieving the unknown qubit which Alice intends to send to Bob. At last Alice is left with no state and the unknown qubit is re-created at Bob's end as a result of Bob's unitary application. The same logic is applicable to the teleportation schemes subjected to higher dimensional systems where instead of considering a single qubit Alice is thought to be in possession with a single qutrit and she plans to send it to Bob via two qutrit entangled resource at the expense of three trits of classical message. In this regard, recently Huang et al teleported a single qubit and a single qutrit from Alice to Bob via two qutrit entangled state as quantum channel\cite{huang2020quantum}. In the ideal situation, the teleportation process involves unitary operators, which preserve the quantum state's norm. However, in real-world scenarios, such as imperfections like noise or decoherence, can ask for participation of non-unitary operations, which alter the quantum state's norm. These non-unitary operators lead to imperfect teleportation, where the teleported state is not an exact copy of the original state, but rather a $``$noisy" version\cite{sala2024quantum}. Moreover, it is not always the case that the optimal state extraction strategy of the unknown state obtained from Alice's end should be aligned with basic teleporation process. This has recently been shown with respect to partially entangled states by Roa et al\cite{roa2019recovering}. In the present work we continue with the scheme of teleportation where Alice holds an unknown single qutrit in her possession and during the process she shares two qutrit entangled channel with Bob to whom she intends to communicate this unknown qutrit. Huang et al\cite{huang2020quantum} had studied the teleportation scheme with a two qutrit state as teleportation channel and later that was experimentally verified by Hu et al \cite{hu2020experimental}. They did the experiment with the two qutrit entangled quantum channel which is singlet state in qutrit system. In this paper we have shown that there is another two qutrit entangled state belonging to $SU(3)$ group which also can be used as quantum channel for single qutrit teleportation scheme.  In this context, we have also made a comparative study of the teleportation protocols by using two different two qutrit entangled resources that Alice will share with Bob. We shall first re-visit the teleportation protocol with singlet type two qutrit state as channel. For the convenience of the readers of this article we designate this singlet resource (or channel) by $\chi^{U}$. Next we consider the other two qutrit entangled resource and denote that by $\chi^{NU}$ and perform the teleportation task. However the notations $\chi^{U}$ and $\chi^{NU}$ have been chosen cleverly here.  Thus the Bennett protocol is run as usual where Alice is made to club her single qutrit to the shared two qutrit channel with an exception here that an an auxiliary basis states\cite{leslie2019maximal}, which we name in this article as Leslie basis, have been involved and with this basis at hand Alice manipulates her shared channel to complete the task of teleportation. Another important fact to be noted here is that when Alice share $\vert \chi^U\rangle$ as channel. Bob needs to apply a set of unitary operators to retrieve the original single qutrit while the set of operators chosen by Bob is non-unitary when the shared entangled resource is $\vert \chi^{NU}\rangle$. The paper is designed as follows.\\\\
In section $II$ we give brief overview of the channels shared by Alice and Bob, denoted as $\chi^{U}$ and $\chi^{NU}$. Section $III$ has been divided into two parts, part $(a)$ and part $(b)$. Part $(a)$ revisits the teleportation scheme with two qutrit entangled state $\chi^{U}$  as quantum channel for doing teleportation but with the assistance of Leslie entangled basis states, and in this sense our approach is different from that was shown in the paper by Huang et al,  while part $(b)$ focuses on our main result where the state $\chi^{NU}$ has been used as quantum channel for the teleportation protocol. This is followed by the conclusion and possible future directions in section $IV$.
\section{Two qutrit states as quantum channels:}
\label{basiststates}
\noindent In this section we introduce the entangled channels that Alice will share with Bob for the successful performance in teleporting her unknown single qutrit to Bob.  Sen and Dey \cite{sen2024inequality} have proposed a set of two qutrit basis states $\mathcal{B}_1$ which is taken into consideration and  from there the two channels are selected, one of which is of singlet form and  we like to denote this by $\chi^{U}$ and another state (we call it as octet state) has been denoted by $\chi^{NU}$, for the purpose of the scheme of teleportation. These two states of basis $\mathcal{B}_I$ can be generated from $SU(3)$ group and they are the normalized two qutrit entangled states. The states will then be used as channels which will be clubbed with Alice's single qutrit state. For the sake of our discussion, we would sometimes refer  these channels suggested by Sen et al as (a) class $1$ type channel (for $\chi^U$) and (b) class $2$ type channel ($\chi^{NU}$). Though class $1$ qutrit state was studied earlier, the efficacy of class $2$ qutrit state has yet to be explored in the context of teleportation. Yet, we have discussed both the states from the perspective of standard single qutrit teleportation from Alice to Bob and for the sole purpose of making comparative arguments. The channels are mathematically expressed as
\begin{eqnarray}
\label{senclass1basisstates}
\vert \chi^{U}_{BC}\rangle &=&  \frac{1}{\sqrt{3}}\Big(\vert 0_{B}\rangle\vert 0_{C}\rangle + \vert 1_{B}\rangle\vert 1_{C}\rangle + \vert 2_{B}\rangle\vert 2_{C}\rangle\Big),
\end{eqnarray}
\begin{eqnarray}
\label{senclass2basisstates}
\vert \chi^{NU}_{BC}\rangle &=& \frac{1}{\sqrt{6}}\Big(-2\vert 0_{B}\rangle\vert 0_{C}\rangle + \vert 1_{B}\rangle\vert 1_{C}\rangle + \vert 2_{B}\rangle\vert 2_{C}\rangle\Big).
\end{eqnarray}
To perform teleportation process with the above two channels, unlike Bell basis, an auxiliary basis will be required which Alice will use to manipulate the joint system for further processing. A basis of states set by Leslie et al may be taken into consideration\cite{leslie2019maximal}.
With implicit symmetrization or anti-symmetrization process such basis is constructed with nine normalized elements in it. We denote that auxiliary basis, defined by Leslie, as $\mathcal{B}_2$. 
\begin{eqnarray}
\label{lesliebasis}
\vert \Psi^{0}_{BC}\rangle &=& \frac{1}{\sqrt{3}}\Big[ \vert 0_{B}\rangle\vert 0_{C}\rangle + \vert 1_{B}\rangle \vert 1_{C}\rangle+ \vert 2_{B}\rangle\vert 2_{c}\rangle\Big],\nonumber\\
\vert \Psi^{1}_{BC}\rangle &=& \frac{1}{\sqrt{3}}\Big [\vert 0_{B}\rangle\vert 0_{C}\rangle +e^{i\frac{2\pi}{3}}\vert 1_{B}\rangle\vert 1_{C}\rangle + e^{i\frac{4\pi}{3}} \vert 2_{B}\rangle\vert 2_{C}\rangle\Big],\nonumber\\
\vert \Psi^{2}_{BC}\rangle &=& \frac{1}{\sqrt{3}}\Big [\vert 0_{B}\rangle\vert 0_{C}\rangle + e^{i\frac{4\pi}{3}} \vert 1_{B}\rangle \vert 1_{C}\rangle + e^{i\frac{2\pi}{3}} \vert 2_{B}\rangle \vert 2_{C}\rangle\Big],\nonumber\\
\vert \Psi^{3}_{BC}\rangle &=& \frac{1}{\sqrt{3}}\Big[ \vert 0_{B}\rangle\vert 1_{C}\rangle + \vert 1_{B}\rangle \vert 2_{C}\rangle + \vert 2\rangle_{B}\vert 0_{C}\rangle\Big],\nonumber\\
\vert \Psi^{4}_{BC}\rangle &=& \frac{1}{\sqrt{3}}\Big [\vert 0_{B}\rangle \vert 1_{C}\rangle + e^{i\frac{2\pi}{3}}\vert 1_{B}\rangle \vert 2_{C}\rangle + e^{i\frac{4\pi}{3}}\vert 2_{B}\rangle \vert 0_{C}\rangle\Big],\nonumber\\
\vert \Psi^{5}_{BC}\rangle &=& \frac{1}{\sqrt{3}}\Big [\vert 0_{B}\rangle \vert 1_{C}\rangle + e^{i\frac{4\pi}{3}}\vert 1_{B}\rangle \vert 2_{C}\rangle + e^{i\frac{2\pi}{3}} \vert 2_{B}\rangle \vert 0_{C}\rangle\Big],\nonumber\\
\vert \Psi^{6}_{BC}\rangle &=& \frac{1}{\sqrt{3}}\Big [\vert 0_{B}\rangle \vert 2_{C}\rangle + \vert 1_{B}\rangle \vert 0_{C}\rangle + \vert 2_{B}\rangle \vert 1_{C}\rangle],\nonumber\\
\vert \Psi^{7}_{BC}\rangle &=& \frac{1}{\sqrt{3}}\Big [\vert 0_{B}\rangle \vert 2_{C}\rangle + e^{i\frac{2\pi}{3}}\vert 1_{B}\rangle \vert 0_{C}\rangle + e^{i\frac{4\pi}{3}} \vert 2_{B}\rangle \vert 1_{C}\rangle\Big],\nonumber\\
\vert \Psi^{8}_{BC}\rangle &=& \frac{1}{\sqrt{3}}\Big [\vert 0_{B}\rangle \vert 2_{C}\rangle + e^{i\frac{4\pi}{3}}\vert 1_{B}\rangle \vert 0_{C}\rangle + e^{i\frac{2\pi}{3}} \vert 2_{B}\rangle \vert 1_{C}\rangle\Big].\nonumber\\
\end{eqnarray}
Before proceeding, to avoid notational ambiguity, we assume that the state with suffix $\vert a_{A}\rangle$ will be the single qutrit state that Alice will intend to send to Bob and the two qutrit states with suffix $\vert b_{B}\rangle\otimes\vert c_{C}\rangle$ will denote the channels shared by Alice and Bob, where suffix $B$ represents the qutrit held by Alice and the suffix $C$ is the qutrit held by Bob from the channels, (where $a=0,1,2$,\:$b=0,1,2$ and $c=0,1,2$). 
\section{Qutrit Teleportation:}
\noindent Now we will discuss the said teleportation protocol which will be considered in our following investigations. 
For any teleportation scheme the sender needs to share a quantum channel with the receiver, which the sender will use to send her state. The sender also takes the help of classical communication (which may be a usual phone call to the receiver and like).  We divide the scheme in two parts, where in each case, the sender Alice holds an unknown single qutrit whose mathematical abstraction is
 \begin{eqnarray}
\label{singlequtri1}
\vert \phi_{A}\rangle = \alpha \vert 0_{A}\rangle + \beta \vert 1_{A}\rangle + \gamma \vert 2_{A}\rangle.
\end{eqnarray}
\subsection*{Part $(a)$:}
\subsubsection*{Derivation:}
\noindent  The state $\vert \chi^{U}_{BC}\rangle$ of Eq.(\ref{senclass1basisstates}) from class $\mathcal{B}_1$ basis states is the quantum channel that Alice shares with Bob. Leslie basis states i.e. $\mathcal{B}_2$ of Eq.(\ref{lesliebasis}) can be re-expressed as follows.
\begin{eqnarray}
\label{senre1}
\vert 0_{B}\rangle \vert 0_{C}\rangle &=& \frac{1}{\sqrt{3}}\Big[ \vert \Psi^{0}_{BC} \rangle+ \vert \Psi^{1}_{BC}\rangle + \vert \Psi^{2}_{BC}\rangle\Big],\nonumber\\
\vert 0_{B}\rangle \vert 1_{C}\rangle &=& \frac{1}{\sqrt{3}}\Big[ \vert \Psi^{3}_{BC}\rangle + \vert \Psi^{4}_{BC}\rangle + \vert \Psi^{5}_{BC}\rangle\Big],\nonumber\\
\vert 0_{B}\rangle \vert 2_{C}\rangle &=& \frac{1}{\sqrt{3}}\Big[ \vert \Psi^{6}_{BC}\rangle + \vert \Psi^{7}_{BC}\rangle + \vert \Psi^{8}_{BC}\rangle\Big],\nonumber\\
\vert 2_{B}\rangle \vert 1_{C}\rangle &=& \frac{1}{\sqrt{3}}\Big[ \vert \Psi^{6}_{BC}\rangle + e^{i\frac{2\pi}{3}}\vert\Psi^{7}_{BC}\rangle + e^{i\frac{4\pi}{3}}\vert\Psi^{8}_{BC}\rangle\Big],\nonumber\\
\vert 1_{B}\rangle \vert 0_{C}\rangle &=& \frac{1}{\sqrt{3}}\Big [\vert \Psi^{6}_{BC}\rangle + e^{i\frac{4\pi}{3}}\vert \Psi^{7}_{BC}\rangle + e^{i\frac{2\pi}{3}} \vert \Psi^{8}_{BC}\rangle\Big],\nonumber\\
\vert 1_{B}\rangle \vert 2_{C}\rangle &=& \frac{1}{\sqrt{3}}\Big [\vert\Psi^{3}_{BC}\rangle +  e^{i\frac{4\pi}{3}}\vert \Psi^{4}_{BC}\rangle +  e^{i\frac{2\pi}{3}} \vert \Psi^{5}_{BC}\rangle\Big],\nonumber\\ 
\vert 2_{B}\rangle \vert 0_{C}\rangle &=& \frac{1}{\sqrt{3}} \Big [\vert\Psi^{3}_{BC}\rangle + e^{i\frac{2\pi}{3}}\vert \Psi^{4}_{BC}\rangle + e^{i\frac{4\pi}{3}}\vert \Psi^{5}_{BC}\rangle\Big]\nonumber\\
\vert 1_{B}\rangle \vert 1_{C}\rangle &=& \frac{1}{\sqrt{3}}\Big [\vert\Psi^{0}_{BC}\rangle +e^{i\frac{4\pi}{3}}\vert \Psi^{1}_{BC} \rangle+ e^{i\frac{2\pi}{3}} \vert\Psi^{2}_{BC}\rangle\Big],\nonumber\\
\vert 2_{B}\rangle \vert 2_{C}\rangle &=& \frac{1}{\sqrt{3}}\Big [\vert\Psi^{0}_{BC}\rangle + e^{i\frac{2\pi}{3}}\vert \Psi^{1}_{BC}\rangle + e^{i\frac{4\pi}{3}} \vert \Psi^{2}_{BC}\rangle\Big],\nonumber\\
\end{eqnarray}
Following the scheme of standard teleportation, Alice clubs the state (\ref{singlequtri1}) with that of the qutrit she holds from the channel $\vert \chi^{U}_{BC}\rangle$ and consequently the following is generated.
\begin{widetext}
\begin{eqnarray}
\label{clubbedtwoqutritchannel1}
\vert \xi^U_{ABC}\rangle &=&\vert \phi_{A}\rangle\otimes \vert \chi^{U}_{BC}\rangle  = \frac{1}{\sqrt{3}}\Big(\vert 0_{A}\rangle\otimes \vert 0_{B}\rangle\otimes \alpha\vert 0_{C}\rangle + \vert 0_{A}\rangle\otimes\vert 1_{B}\rangle\otimes\alpha\vert 1\rangle_{C} + \vert 0_{A}\rangle\otimes \vert 2_{B}\rangle\otimes \alpha\vert 2_{C}\rangle + \vert 1_{A}\rangle\otimes \vert 0_{B}\rangle\otimes \beta\vert 0\rangle_{C}\nonumber\\&& + \vert 1_{A}\rangle\otimes \vert 1_{B}\rangle\otimes \beta\vert 1_{C} \rangle+ \vert 1_{A}\rangle\otimes \vert 2_{B}\rangle\otimes\beta\vert 2_{C}\rangle + \vert 2_{A}\rangle\otimes \vert 0_{B}\rangle\otimes \gamma\vert 0_{C}\rangle  +\vert 2_{A}\rangle\otimes \vert 1_{B}\rangle\otimes \gamma\vert 1\rangle_{C} \nonumber\\&& + \vert 2_{A}\rangle\otimes \vert 2_{B}\rangle\otimes \gamma\vert 2_{C}\rangle\Big).
\end{eqnarray}
\end{widetext}
Alice is now in possession of two qutrits formed by the elements of the type $\vert a_{A}\rangle\otimes\vert b_{B}\rangle$ ($a=0,1,2\:,b=0,1,2$), while Bob holds single qutrit $\vert c_{C}\rangle$ ($c=0,1,2$). Using the relation obtained in (\ref{senre1}), the state (\ref{clubbedtwoqutritchannel1}) can be re-written as 
\begin{widetext}
\begin{eqnarray}
\label{xiabc1}
\vert \xi^U_{ABC}\rangle &=& \frac{1}{3}\Big[\vert \Psi^{0}_{AB}\rangle\otimes \vert s^{0}_{C}\rangle + \vert \Psi^{1}_{AB}\rangle \otimes \vert s^{1}_{C}\rangle  + \vert \Psi^{2}_{AB}\rangle \otimes \vert s^{2}_{C}\rangle + \vert \Psi^{3}_{AB}\rangle \otimes \vert s^{3}_{C}\rangle\nonumber\\&& + \vert \Psi^{4}_{AB}\rangle \otimes \vert s^{4}_{C}\rangle + \vert \Psi^{5}_{AB}\rangle \otimes \vert s^{5}_{C}\rangle + \vert \Psi^{6}_{AB}\rangle  \otimes \vert s^{6}_{C}\rangle + \vert \Psi^{7}_{AB}\rangle \otimes \vert s^{7}_{C}\rangle + \vert \Psi^{8}_{AB}\rangle\otimes \vert s^{8}_{C}\rangle \Big],\nonumber\\
\end{eqnarray}
\end{widetext}
where we have
\begin{eqnarray}
\label{statetobeteleported1}
\vert s^{0}_{C}\rangle &=& \alpha \vert 0_{C}\rangle + \beta \vert 1_{C}\rangle + \gamma \vert 2_{C}\rangle,\nonumber\\
\vert s^{1}_{C}\rangle &=& \alpha \vert 0_{C}\rangle + e^{i\frac{4\pi}{3}}\beta \vert 1_{C}\rangle + e^{i\frac{2\pi}{3}}\gamma \vert 2_{C}\rangle,\nonumber\\
\vert s^{2}_{C}\rangle &=& \alpha \vert 0_{C}\rangle + e^{i\frac{2\pi}{3}}\beta \vert 1_{C}\rangle+ e^{i\frac{4\pi}{3}}\gamma \vert 2_{C}\rangle,\nonumber\\
\vert s^{3}_{C}\rangle &=& \gamma \vert 0_{C}\rangle +\alpha \vert 1_{C}\rangle + \beta \vert 2_{C}\rangle,\nonumber\\
\vert s^{4}_{C}\rangle &=& e^{i\frac{2\pi}{3}}\gamma \vert 0_{C}\rangle + \alpha \vert 1_{C}\rangle + e^{i\frac{4\pi}{3}} \beta \vert 2_{C}\rangle,\nonumber\\
\vert s^{5}_{C}\rangle &=& e^{i\frac{4\pi}{3}}\gamma \vert 0_{C}\rangle +\alpha \vert 1_{C}\rangle + e^{i\frac{2\pi}{3}} \beta \vert 2_{C}\rangle\nonumber\\
\vert s^{6}_{C}\rangle &=&  \beta \vert 0_{C}\rangle + \gamma \vert 1_{C}\rangle + \alpha \vert 2_{C}\rangle,\nonumber\\
\vert s^{7}_{C}\rangle &=& e^{i\frac{4\pi}{3}} \beta \vert 0_{C}\rangle + e^{i\frac{2\pi}{3}}\gamma \vert 1_{C}\rangle + \alpha \vert 2_{C}\rangle,\nonumber\\
\vert s^{8}_{C}\rangle &=& e^{i\frac{2\pi}{3}} \beta \vert 0_{C}\rangle + e^{i\frac{4\pi}{3}}\gamma \vert 1_{C}\rangle + \alpha \vert 2_{C}\rangle.
\end{eqnarray}
Alice will now make joint measurements on her part of two qutrits and informs her measurement outcomes to Bob. This information will be conveyed to Bob via a simple telephone call or by sending email, thus involving classical communication. Suppose the outcome obtained by Alice is $\vert \psi^{0}_{AB}\rangle$  and she informs about her outcome to Bob classically, then Bob needs to apply unitary operation on his counterpart $\vert s^{0}_{C}\rangle$ to retrieve the original qutrit state at his end ($\vert \phi_{C} \rangle$, the perfect replica of $\vert \phi_{A}\rangle$), which was initially held by Alice.  The details of the above process for all the other outcomes at Alice's end is depicted as follows. 
\begin{eqnarray}
\label{xiabc1}
\vert \xi^U_{ABC}\rangle &=& \frac{1}{3}\Big[\vert \Psi^{0}_{AB}\rangle\otimes U_0\vert s^{0}_{C}\rangle + \vert \Psi^{1}_{AB}\rangle \otimes U_1\vert s^{1}_{C}\rangle  + \vert \Psi^{2}_{AB}\rangle \otimes U_2\vert s^{2}_{C}\rangle + \vert \Psi^{3}_{AB}\rangle \otimes U_3\vert s^{3}_{C}\rangle\nonumber\\&& + \vert \Psi^{4}_{AB}\rangle \otimes U_4\vert s^{4}_{C}\rangle + \vert \Psi^{5}_{AB}\rangle \otimes U_5\vert s^{5}_{C}\rangle + \vert \Psi^{6}_{AB}\rangle  \otimes U_6\vert s^{6}_{C}\rangle + \vert \Psi^{7}_{AB}\rangle \otimes U_7\vert s^{7}_{C}\rangle + \vert \Psi^{8}_{AB}\rangle\otimes U_8\vert s^{8}_{C}\rangle \Big],\nonumber\\
\end{eqnarray}
\subsubsection*{Result 1:}
\noindent In this way, to retrieve the single qutrit from Alice, Bob has to depend on the classical message he receives from her and thus we get the following which represent how Bob retrieves them.
\begin{eqnarray}
\label{retreived1}
U_0 \mapsto \langle \Psi^{0}_{AB}\vert U_0 \vert \xi_{ABC}^U\rangle = U_{0}\vert s^{0}_{C}\rangle = \vert \phi_C\rangle, \nonumber\\
U_1 \mapsto \langle \Psi^{1}_{AB}\vert U_1 \vert \xi_{ABC}^U\rangle = U_{1}\vert s^{1}_{C}\rangle = \vert \phi_C\rangle, \nonumber\\
U_2 \mapsto \langle \Psi^{2}_{AB}\vert U_2 \vert \xi_{ABC}^U\rangle = U_{2}\vert s^{2}_{C}\rangle = \vert \phi_C\rangle, \nonumber\\
U_3 \mapsto \langle \Psi^{3}_{AB}\vert U_3 \vert \xi_{ABC}^U\rangle = U_{3}\vert s^{3}_{C}\rangle = \vert \phi_C\rangle, \nonumber\\
U_4 \mapsto \langle \Psi^{4}_{AB}\vert U_4 \vert \xi_{ABC}^U\rangle = U_{4}\vert s^{4}_{C}\rangle = \vert \phi_C\rangle, \nonumber\\
U_5 \mapsto \langle \Psi^{5}_{AB}\vert U_5 \vert \xi_{ABC}^U\rangle = U_{5}\vert s^{5}_{C}\rangle = \vert \phi_C\rangle, \nonumber\\
U_6 \mapsto \langle \Psi^{6}_{AB}\vert U_6 \vert \xi_{ABC}^U\rangle = U_{6}\vert s^{6}_{C}\rangle = \vert \phi_C\rangle, \nonumber\\
U_7 \mapsto \langle \Psi^{7}_{AB}\vert U_7 \vert \xi_{ABC}^U\rangle = U_{7}\vert s^{7}_{C}\rangle = \vert \phi_C\rangle, \nonumber\\
U_8 \mapsto \langle \Psi^{8}_{AB}\vert U_8 \vert \xi_{ABC}^U\rangle = U_{8}\vert s^{8}_{C}\rangle = \vert \phi_C\rangle, \nonumber\\
\end{eqnarray}
where,\\
\begin{eqnarray}
\label{Bob unitaries1}
U_{0} &=& \mathcal{I} = \vert 0_{C}\rangle\langle 0_{C}\vert + \vert 1_{C}\rangle\langle 1_{C}\vert + \vert 2_{C}\rangle\langle 2_{C}\vert,\nonumber\\
U_{1} &=& \vert 0_{C}\rangle\langle 0_{C}\vert  + e^{i\frac{2\pi}{3}}\vert 1_{C}\rangle\langle 1_{C}\vert  + e^{i\frac{4\pi}{3}}\vert 2_{C}\rangle\langle 2_{C}\vert,\nonumber\\
U_{2} &=& \vert 0_{C}\rangle\langle 0_{C}\vert  + e^{i\frac{4\pi}{3}}\vert 1_{C}\rangle\langle 1_{C}\vert  + e^{i\frac{2\pi}{3}}\vert 2_{C}\rangle\langle 2_{C}\vert,\nonumber\\
U_{3} &=& \vert 0_{C}\rangle\langle 1_{C}\vert + \vert 1_{C}\rangle\langle 2_{C}\vert + \vert 2_{C}\rangle\langle 0_{C}\vert,\nonumber\\
U_{4} &=& \vert 0_{C}\rangle\langle 1_{C}\vert  + e^{i\frac{2\pi}{3}}\vert 1_{C}\rangle\langle 2_{C}\vert  + e^{i\frac{4\pi}{3}}\vert 2_{C}\rangle\langle 0_{C}\vert,\nonumber\\
U_{5} &=& \vert 0_{C}\rangle\langle 1_{C}\vert  + e^{i\frac{4\pi}{3}}\vert 1_{C}\rangle\langle 2_{C}\vert  + e^{i\frac{2\pi}{3}}\vert 2_{C}\rangle\langle 0_{C}\vert,\nonumber\\
U_{6} &=& \vert 0_{C}\rangle\langle 2_{C}\vert + \vert 1_{C}\rangle\langle 0_{C}\vert + \vert 2_{C}\rangle\langle 1_{C}\vert,\nonumber\\
U_{7} &=& \vert 0_{C}\rangle\langle 2_{C}\vert  + e^{i\frac{2\pi}{3}}\vert 1_{C}\rangle\langle 0_{C}\vert  + e^{i\frac{4\pi}{3}}\vert 2_{C}\rangle\langle 1_{C}\vert,\nonumber\\
U_{8} &=& \vert 0_{C}\rangle\langle 2_{C}\vert  + e^{i\frac{4\pi}{3}}\vert 1_{C}\rangle\langle 0_{C}\vert  + e^{i\frac{2\pi}{3}}\vert 2_{C}\rangle\langle 1_{C}\vert.
\end{eqnarray}
Hence using the entangled two qutrit states as resource, Alice is finally able to teleport the quantum single qutrit state (\ref{singlequtri1}) at Bob's location (denoted by $\vert \phi_C\rangle$).\\\\
\subsection*{Part $(b)$:}
\subsubsection*{Derivation:}
\noindent We shall now consider the two qutrit entangled channel of Eq.(\ref{senclass2basisstates}), which also belongs to $SU(3)$ group. As before, Alice holds a single qutrit as defined in eq.(\ref{singlequtri1}).  She wants to communicate this unknown state to the recipient Bob. 
In this channel (\ref{senclass2basisstates}), the first qutrit is held by Alice and the second qutrit is held by Bob. Following the Bennett's teleportation protocol once again,  Alice clubs her single qutrit state (\ref{singlequtri1}) to the channel (\ref{senclass2basisstates}) and consequently the following is generated.
\begin{widetext}
\begin{eqnarray}
\label{clubbedtwoqutritchannel2}
\vert \xi^{NU}_{ABC}\rangle &=&\vert \phi_{A}\rangle\otimes \vert\chi^{NU}_{BC}\rangle = \frac{1}{\sqrt{6}}\Big(-2\vert 0_{A}\rangle\otimes \vert 0_{B}\rangle\otimes \alpha\vert 0\rangle_{C} + \vert 0_{A}\rangle \otimes 1_{B}\rangle\otimes \alpha\vert 1_{C}\rangle + \vert 0_{A}\rangle\otimes  \vert 2_{B}\rangle \otimes \alpha\vert 2_{C}\rangle -2\vert 1_{A}\rangle\otimes \vert 0_{B}\rangle\otimes \beta\vert 0_{C}\rangle \nonumber\\ && + \vert 1_{A}\rangle \otimes \vert 1_{B}\rangle\otimes \beta\vert 1_{C}\rangle  + \vert 1_{A}\rangle\otimes  \vert 2_{B}\rangle\otimes \beta\vert 2_{C}\rangle + -2\vert 2_{A}\rangle \vert 0_{B}\rangle \otimes \gamma\vert 0_{C}\rangle +\vert 2_{A}\rangle \vert 1_{B}\rangle\otimes \gamma\vert 1_{C}\rangle + \vert 2_{A}\rangle \otimes  \vert 2_{B}\rangle \otimes \gamma\vert 2_{C}\rangle\Big).\nonumber\\
\end{eqnarray}
\end{widetext}
Alice again holds two qutrits, while Bob holds one (as shown in eq.(\ref{clubbedtwoqutritchannel2})). Using the basis states eq.(\ref{lesliebasis}), the state (\ref{clubbedtwoqutritchannel2}) can be re-written as 
\begin{widetext}
\begin{eqnarray}
\label{clubbedtwoqutritchannel12}
\vert \xi^{NU}_{ABC}\rangle  &=& \frac{1}{\sqrt{18}}\Big[\vert \Psi^{0}_{AB}\rangle\otimes  \vert q^{0}_{C}\rangle + \vert \Psi^{1}_{AB}\rangle\otimes  \vert q^{1}_{C}\rangle + \vert \Psi^{2}_{AB}\rangle\otimes \vert q^{2}_{C}\rangle  + \vert \Psi^{3}_{AB}\rangle\otimes  \vert q^{3}_{C}\rangle + \vert \Psi^{4}_{AB}\rangle\otimes \vert q^{4}_{C}\rangle \nonumber\\ &&+ \vert \Psi^{5}_{AB}\rangle\otimes \vert q^{5}_{C}\rangle + \vert \Psi^{6}_{AB}\rangle\otimes \vert q^{6}_{C}\rangle + \vert \Psi^{7}_{AB}\rangle\otimes\vert \vert q^{7}_{C}\rangle + \vert \Psi^{8}_{AB}\rangle\otimes \vert q^{8}_{C}\rangle \Big]\nonumber\\,
\end{eqnarray}
\end{widetext}
Then following Bennett's protocol as described for part $(a)$, but using non-unitary operators in this case, the Eq.(\ref{clubbedtwoqutritchannel12}) can be written as
\begin{eqnarray}
\label{xiabc2b}
\vert \xi^{NU}_{ABC}\rangle &=& \frac{1}{3}\Big[\vert \Psi^{0}_{AB}\rangle\otimes NU_0\vert q^{0}_{C}\rangle + \vert \Psi^{1}_{AB}\rangle \otimes NU_1\vert q^{1}_{C}\rangle  + \vert \Psi^{2}_{AB}\rangle \otimes NU_2\vert q^{2}_{C}\rangle + \vert \Psi^{3}_{AB}\rangle \otimes NU_3\vert q^{3}_{C}\rangle\nonumber\\&& + \vert \Psi^{4}_{AB}\rangle \otimes NU_4\vert q^{4}_{C}\rangle + \vert \Psi^{5}_{AB}\rangle \otimes NU_5\vert q^{5}_{C}\rangle + \vert \Psi^{6}_{AB}\rangle  \otimes NU_6\vert q^{6}_{C}\rangle + \vert \Psi^{7}_{AB}\rangle \otimes NU_7\vert q^{7}_{C}\rangle + \vert \Psi^{8}_{AB}\rangle\otimes NU_8\vert q^{8}_{C}\rangle \Big],\nonumber\\
\end{eqnarray}
where we have
\begin{widetext}
\begin{eqnarray}
\label{statetobeteleported1}
\vert q^{0}_{C}\rangle &=& -2\alpha \vert 0_{C}\rangle + \beta \vert 1_{C}\rangle + \gamma \vert 2_{C}\rangle,\nonumber\\
\vert q^{1}_{C}\rangle &=& -2\alpha \vert 0_{C}\rangle +  e^{i\frac{4\pi}{3}}\beta \vert 1_{C}\rangle +  e^{i\frac{2\pi}{3}}\gamma \vert 2_{C}\rangle,\nonumber\\
\vert q^{2}_{C}\rangle &=&  -2\alpha \vert 0_{C}\rangle +  e^{i\frac{2\pi}{3}}\beta \vert 1_{C}\rangle +  e^{i\frac{4\pi}{3}}\gamma \vert 2_{C}\rangle,\nonumber\\
\vert q^{3}_{C}\rangle &=& 2 \gamma \vert 0_{C}\rangle + \alpha \vert 1_{C}\rangle + \beta \vert 2_{C}\rangle,\nonumber\\
\vert q^{4}_{C}\rangle &=&  -  2 e^{i\frac{2\pi}{3}}\gamma \vert 0_{C}\rangle + \alpha \vert 1_{C}\rangle +  \frac{e^{i\frac{4\pi}{3}}}{2}\beta \vert 2_{C}\rangle\nonumber\\
\vert q^{5}_{C}\rangle &=& -2 e^{i\frac{4\pi}{3}}\gamma \vert 0_{C}\rangle + \alpha \vert 1_{C}\rangle +e^{i\frac{2\pi}{3}}\beta \vert 2_{C}\rangle\nonumber\\
\vert q^{6}_{C}\rangle &=& -2\beta \vert 0_{C}\rangle + \gamma \vert 1_{C}\rangle + \alpha \vert 2_{C}\rangle\nonumber\\
\vert q^{7}_{C}\rangle &=&  - 2e^{i\frac{4\pi}{3}} \beta \vert 0_{C}\rangle+ e^{i\frac{2\pi}{3}}\gamma \vert 1_{C}\rangle + \alpha \vert 2_{C}\rangle\nonumber\\
\vert q^{8}_{C}\rangle &=& -2e^{i\frac{2\pi}{3}} \beta \vert 0_{C}\rangle + e^{i\frac{4\pi}{3}}\gamma \vert 1_{C}\rangle + \alpha \vert 2_{C}\rangle.
\end{eqnarray}
\end{widetext}
\subsubsection*{Result 2:}
\noindent The following table summarizes the measurement outcomes of Alice and her communication to Bob and subsequently the non-unitary operators which Bob applies, will retrieve the single qutrit state $\vert \phi_{A}\rangle$ of Eq.(\ref{singlequtri1}) that Alice wants to send to Bob, at Bob's end (we symbolize the retrieved state as $\vert \phi_{C}\rangle$). As before, depending on the classical message from Alice, Bob can retrieve the original single qutrit of Eq.(\ref{singlequtri1})  by applying the following operators which are non-unitary.
\begin{eqnarray}
\label{retreived2}
NU_0 \mapsto \langle \Psi^{0}_{AB}\vert NU_0 \vert \xi_{ABC}^U\rangle = NU_{0}\vert q^{0}_{C}\rangle = \vert \phi_C\rangle, \nonumber\\
NU_1 \mapsto \langle \Psi^{1}_{AB}\vert NU_1 \vert \xi_{ABC}^U\rangle = NU_{1}\vert q^{1}_{C}\rangle = \vert \phi_C\rangle, \nonumber\\
NU_2 \mapsto \langle \Psi^{2}_{AB}\vert NU_2 \vert \xi_{ABC}^U\rangle = NU_{2}\vert q^{2}_{C}\rangle = \vert \phi_C\rangle, \nonumber\\
NU_3 \mapsto \langle \Psi^{3}_{AB}\vert NU_3 \vert \xi_{ABC}^U\rangle = NU_{3}\vert q^{3}_{C}\rangle = \vert \phi_C\rangle, \nonumber\\
NU_4 \mapsto \langle \Psi^{4}_{AB}\vert NU_4 \vert \xi_{ABC}^U\rangle = NU_{4}\vert q^{4}_{C}\rangle = \vert \phi_C\rangle, \nonumber\\
NU_5 \mapsto \langle \Psi^{5}_{AB}\vert NU_5 \vert \xi_{ABC}^U\rangle = NU_{5}\vert q^{5}_{C}\rangle = \vert \phi_C\rangle, \nonumber\\
NU_6 \mapsto \langle \Psi^{6}_{AB}\vert NU_6 \vert \xi_{ABC}^U\rangle = NU_{6}\vert q^{6}_{C}\rangle = \vert \phi_C\rangle, \nonumber\\
NU_7 \mapsto \langle \Psi^{7}_{AB}\vert NU_7 \vert \xi_{ABC}^U\rangle = NU_{7}\vert q^{7}_{C}\rangle = \vert \phi_C\rangle, \nonumber\\
NU_8 \mapsto \langle \Psi^{8}_{AB}\vert NU_8 \vert \xi_{ABC}^U\rangle = NU_{8}\vert q^{8}_{C}\rangle = \vert \phi_C\rangle, \nonumber\\
\end{eqnarray}
where,
\begin{eqnarray}
\label{Bob unitaries1}
NU_{0} &=& -\frac{1}{2}\vert 0_{C}\rangle\langle 0_{C}\vert + \vert 1_{C}\rangle\langle 1_{C}\vert + \vert 2_{C}\rangle\langle 2_{C}\vert,\nonumber\\
NU_{1} &=& -\vert 0_{C}\rangle\langle 0_{C}\vert  + 2e^{i\frac{4\pi}{3}}\vert 1_{C}\rangle\langle 1_{C}\vert  + e^{i\frac{2\pi}{3}}\vert 2_{C}\rangle\langle 2_{C}\vert,\nonumber\\
NU_{2} &=& \frac{1}{2}\vert 0_{C}\rangle\langle 0_{C}\vert  + e^{i\frac{2\pi}{3}}\vert 1_{C}\rangle\langle 1_{C}\vert  + e^{i\frac{4\pi}{3}}\vert 2_{C}\rangle\langle 2_{C}\vert,\nonumber\\
NU_{3} &=& \vert 0_{C}\rangle\langle 1_{C}\vert + \vert 1_{C}\rangle\langle 2_{C}\vert - \frac{1}{2}\vert 2_{C}\rangle\langle 0_{C}\vert,\nonumber\\
NU_{4} &=& 2\vert 0_{C}\rangle\langle 1_{C}\vert  + 2 e^{i\frac{4\pi}{3}}\vert 1_{C}\rangle\langle 2_{C}\vert  - e^{i\frac{2\pi}{3}}\vert 2_{C}\rangle\langle 0_{C}\vert,\nonumber\\
NU_{5} &=& \vert 0_{C}\rangle\langle 1_{C}\vert  + e^{i\frac{2\pi}{3}}\vert 1_{C}\rangle\langle 2_{C}\vert -\frac{1}{2}e^{i\frac{4\pi}{3}}\vert 2_{C}\rangle\langle 0_{C}\vert,\nonumber\\
NU_{6} &=& \vert 0_{C}\rangle\langle 2_{C}\vert -\frac{1}{2} \vert 1_{C}\rangle\langle 0_{C}\vert + \vert 2_{C}\rangle\langle 1_{C}\vert,\nonumber\\
NU_{7} &=& 2\vert 0_{C}\rangle\langle 2_{C}\vert  - e^{i\frac{4\pi}{3}}\vert 1_{C}\rangle\langle 0_{C}\vert  + 2 e^{i\frac{4\pi}{3}}\vert 2_{C}\rangle\langle 1_{C}\vert,\nonumber\\
NU_{8} &=& \vert 0_{C}\rangle\langle 2_{C}\vert  -\frac{1}{2} e^{i\frac{2\pi}{3}}\vert 1_{C}\rangle\langle 0_{C}\vert  + e^{i\frac{4\pi}{3}}\vert 2_{C}\rangle\langle 1_{C}\vert.
\end{eqnarray}
\\
Again it has been established that, using the entangled two qutrit state (\ref{senclass2basisstates}) as resource, Alice is finally able to teleport a quantum single qutrit state (\ref{singlequtri1}) to Bob.
\section{Entanglement and Teleportation fidelity of the two-qutrit states:}
\noindent We shall now discuss how efficient the channels described in Eq.(\ref{senclass1basisstates}) and Eq.(\ref{senclass2basisstates}) are in teleporting the single qutrit (\ref{singlequtri1}). We also calculate the nature of the entanglement of these states and to study this we will resort to calculating the von-Neumann entropy (which is sometimes also known as \textit{entanglement entropy}).
\subsection{Entropy of Entanglement of the qutrit states:}
\noindent The von Neumann entropy of a quantum state $\rho$ is given by the formula,
\begin{eqnarray}
\label{entropy}
S(\rho) = -tr(\rho\:\log_{3}\:\rho).
\end{eqnarray}
If $\lambda_{i}$ are the eigenvalues of $\rho$, then the von-Neumann entropy can be computationally expressed as 
\begin{eqnarray}
\label{entropycompute}
S(\rho) = -\sum_{i}\lambda_{i}\log(\lambda_{i}).
\end{eqnarray}
Given the two qutrit state (\ref{senclass1basisstates})(i.e $\vert \chi^{U}\rangle$), we partially trace out either the party $B$ (or  $C$) and using Eq.(\ref{entropycompute}), we find that
\begin{eqnarray}
\label{entropyu}
S(\rho^{\chi^U}_{B}) = S(\rho^{\chi^U}_{C}) = 1.585,
\end{eqnarray}
where $\rho^{\chi^U}$ is density matrix corresponding to the state $\vert \chi^{U}\rangle$. 
On the other hand, given the state (\ref{senclass2basisstates}) (i.e. $\vert \chi^{NU}\rangle$), denoting the corresponding density matrix by $\rho^{\chi^{NU}}$, similarly we observe that
\begin{eqnarray}
\label{entropynu}
S(\rho^{\chi^{NU}_{B}}) = S(\rho^{\chi^{NU}_{A}}) = 1.252
\end{eqnarray}
In qutrit system (i.e.$3\times 3$ system), where each subsystem is defined with respect to the basis $\lbrace \vert 0\rangle, \vert 1\rangle, \vert 2\rangle\rbrace$, for a bipartite pure state $\vert \psi\rangle$, maximum possible entropy $S_{max} = \log_{2}(\min(3,3)) = \log_{2}(3) \approx 1.585$. Thus a state is maximally entangled if its entropy of entanglement reaches this maximum value and we see that this is true for the state $\vert \chi^{U}\rangle$ whose von Neumann entropy is $1.585$. However, when a state is not maximally entangled then this means that the eigenvalues of the reduced system are not equal to all $\frac{1}{3}$. Since $1 < \log_{2}(3) \approx 1.585$, there is range of entropies for entangled states that are not maximally entangled but still have entropy greater than $1$. The state $\vert \chi^{NU}\rangle$ falls into that category, while the von Neumann entropy of $\vert \chi^{NU}\rangle$ is approximately $1.252< \log_{2}(3) $.
\subsection{Teleportation fidelity of the qutrit states:}
\noindent Negativity is a metric that quantifies the level of entanglement between two qubits, providing a numerical value for their quantum correlation. For a density matrix $\rho$ associated with a state, the negativity is denoted by $N(\rho)$ and the expression for evaluation of negativity is \textcolor{blue}{\cite{vidal2002}}
\begin{eqnarray}
\label{negativity}
N_{\rho} &=& \frac{1}{2}\Big(\|\rho^{T_{A}}\|_{1} - 1\Big) = \vert \sum_{i} \lambda_{i}\vert = \frac{1}{2}\sum_{j}\Big[\Big(\vert \lambda_{j}\vert - \lambda_{j}\Big)\Big].
\end{eqnarray}
where, $\rho^{T_{A}}$ is the partial transpose of the composite system $\rho^{AB}$, (taken with respect to first party $A$). Here in above expression subscript $i$ runs over the subset of negative eigenvalues of $\rho^{T_{A}}$ whereas subscript $j$ runs over all the eigenvalues of $\rho^{T_{A}}$. The teleportation fidelity of the state $\rho$
\begin{eqnarray}
\label{telepfid}
f_{\rho} &=& \frac{1 + N_{\rho}}{2}.
\end{eqnarray}
The two states $\vert \chi^{U}\rangle$ and $\vert \chi^{NU}\rangle$ are normalized entangled qutrit states belonging to $SU(3)$ group as proposed by Dey and Sen \cite{sen2024inequality}. Using Eq.(\ref{negativity}) and (\ref{telepfid}) we calculate the teleportation fidelities of the states (\ref{senclass1basisstates}) and (\ref{senclass2basisstates}). It is observed that for the singlet state $\vert \chi^{U}\rangle$, the teleportation fidelity is $1$ while for the state $\vert \chi^{NU}\rangle$, the teleportation fidelity is $\frac{11}{12}\approx 0.9$. Thus we observe that if the non-unitary operator applied by the recipient Bob, the teleportation will be imperfect and consequently the quantum information is degraded.
To above findings can be summarized in the following table where $EoE$ and $TP$ respectively denote the entropy of entanglement and teleportation fidelity respectively:
\begin{table}[h!]
\begin{center}
\caption{Two qutrit channels, their entanglement and teleportation fidelity}
\label{table1}
\begin{tabular}{|c|c|c|}
\hline
\textbf{State} & $EoE$ & $TP$\\
\hline
$\vert \chi^{U}\rangle$ & $1.585$  & $1$\\
\hline
$\vert \chi^{NU}\rangle$ & $1.252$ & $0.9$ \\
\hline
\end{tabular}
\end{center}
\end{table}
\section{Conclusion:} \noindent We have elevated the Bennett scheme of teleportation in the qutrit system. In this context, we have shown that there exist two qutrit entangled channels $\vert \chi^U\rangle$ and $\vert \chi^{NU}\rangle$ belonging to class $\mathcal{B}_1$, which can be viable candidates as quantum channels used for the purposeof of teleportation. The observations that have been made from our studies are (a) the channel $\vert \chi^U\rangle$, when used as quantum channel, Bob is able to extract the unknown quantum single qutrit state unitarily by deciphering the classical message received from Alice, whereas (b) for the channel $\vert \chi^{NU}\rangle$ state as quantum channel, Bob's extraction process is non-unitary. One can put the second case in the frame that, Bob's extraction process is subject to imperfections. In another words, Bob is receiving non-unitarily, the corrupted cousin of Alice's single qutrit. Another difference that exists between the protocol described here and that of Bennett's usual methodology is that Bennett applied Bell basis as shared entangled resource and used those basis states for the recreation of original qutrit's replica while in our case Alice and Bob share two different entangled resources and use another class of basis states $\mathcal{B}_2$ as auxiliary basis, which would help Alice and Bob to perform the protocol of teleportation. Thus, we have shown that Alice, holding a single qutrit, shares two qutrit entangled states of class $\mathcal{B}_{1}$ with Bob. Taking help from an auxiliary basis state $\mathcal{B}_{2}$, Alice is able to perform perfect and imperfect teleportation depending on whether Bob selects unitary or non-unitary operators, respectively, to retrieve the original unknown single qutrit. In future one can study these channels in more depth and try to look for some generalized relations that may connect teleportation fidelities and amount of entanglement in qutrit system. Moreover, the qutrit states can also be tested for the purpose of dense coding which is considered to be the reverse of the teleportation protocol.
\noindent 
\vskip 0.5cm
{\bf \noindent Declaration of competing interest} The authors declare that they have no known competing financial interests or 
personal relationships that could have appeared to influence the work reported in this paper.
\vskip 0.5cm
{\bf \noindent Data availability statement} All data that support the findings of this study are included within the article. No supplementary file has been added.
\bibliographystyle{apsrev}

\bibliography{RefsTeleport}

\end{document}